# A "Toy" Model for Operational Risk Quantification using Credibility Theory

**Hans Bühlmann**
ETH Zürich, Department of Mathematics, HG J58, CH-8092, Zurich, Switzerland.
e-mail: hbuhl@math.ethz.ch

**Pavel V. Shevchenko**
CSIRO Mathematical and Information Sciences, Sydney, Locked Bag 17, North Ryde, NSW, 1670, Australia. e-mail: Pavel.Shevchenko@csiro.au

**Mario V. Wüthrich**
ETH Zürich, Department of Mathematics, HG G 32.5, CH-8092, Zurich, Switzerland.
e-mail: wueth@math.ethz.ch

**Abstract**
To meet the Basel II regulatory requirements for the Advanced Measurement Approaches in operational risk, the bank's internal model should make use of the internal data, relevant external data, scenario analysis and factors reflecting the business environment and internal control systems. One of the unresolved challenges in operational risk is combining of these data sources appropriately. In this paper we focus on quantification of the low frequency high impact losses exceeding some high threshold. We suggest a full credibility theory approach to estimate frequency and severity distributions of these losses by taking into account bank internal data, expert opinions and industry data.

# 1 Introduction

Under the Basel II requirements, BIS (2005), a bank intending to use the Advanced Measurement Approaches (AMA) for quantification of operational risk losses should demonstrate the accuracy of the internal model within Basel II risk cells relevant to the bank. The industry usually refers these risk cells as "risk nodes" in the regulatory matrix of eight business lines times seven risk types. To meet regulatory requirements, the model should make use of the internal data, relevant external data, scenario analysis and factors reflecting business environment and internal control systems. There are various aspects of operational risk modelling, see for example Chavez-Demoulin, Embrechts and Nešlehová (2006), Cruz (2004). Under the Loss Distribution Approach (LDA) of AMA, the banks quantify distributions for the frequency and severity of operational losses for each risk cell over a one year time horizon. The banks can use their own risk cell structure but must be able to map the losses to the Basel II risk cells. The commonly used LDA model for an annual loss $Z$ in a risk cell is a compound process

$$Z = \sum_{k=1}^{N} X_k, \qquad (1)$$

where $N$ is the annual number of events modelled as a random variable from some discrete distribution $P(.|\boldsymbol{\theta})$ and $X_k, k=1,...,N$, are the severities of the events modelled as independent random variables from a continuous distribution $F(.|\boldsymbol{\xi})$. Here, $\boldsymbol{\theta}$ and $\boldsymbol{\xi}$ are distribution parameters (risk profile latent variables). The frequency $N$ and the severities $X_k, k=1,...,N$, are assumed conditionally (given $\boldsymbol{\theta}$ and $\boldsymbol{\xi}$) independent. In our approach $\boldsymbol{\theta}$ and $\boldsymbol{\xi}$ are modelled as realizations of random variables drawn from some distribution. We will consider a group of risks where $\boldsymbol{\theta}$ and $\boldsymbol{\xi}$ are different for the different risks but are drawn from distribution common across the risks. In the proposed framework we do not consider risks individually but regard each risk as embedded in a group of "similar" risks. Constructing such risk groups is important in practice but will not be discussed here.

In this paper we focus on quantification of the low frequency high impact losses exceeding some high threshold for a risk cell (often, most of the operational risk capital is due to these losses). Estimation of the frequency and severity distributions of such losses for each risk cell is a challenging task. The bank's internal data are typically collected over several years and contain few low frequency high impact losses. The industry data are available through external databases from vendors (e.g. OpVar® Database) and consortia of banks (e.g. ORX). Typically, vendors provide data for losses above US$1million while consortium-based data collection threshold is of the order of US$10,000. These are difficult to use directly due to different volumes and other factors. Moreover, the data, probably, have a survival bias. Scenario analysis is undertaken by banks to obtain some quantitative assessment of the risk frequency and severity distributions using expert opinions, see for example Alderweireld, Garcia and Léonard (2006). By itself, scenario analysis is very subjective and should be combined with the analysis of the actual loss data. Combining of internal data with external data and expert opinions in appropriate way is one of the unresolved challenges in operational risk. In practice, ad-hoc procedures are often used to combine these data sources for estimation of the frequency and severity distributions in a risk cell. For example: a) the severity distribution is fitted to the combined samples of internal and external data; b) estimated event arrival rates $\lambda_{ext}$ and $\lambda_{int}$ (implied by external and internal data) are combined as $w\lambda_{int} + (1-w)\lambda_{ext}$ using expert specified (or ad-hoc calculated) weight $w$ to estimate frequency distribution; c) the severity distributions $F_{SA}(X)$, $F_{int}(X)$ and $F_{ext}(X)$ are fitted (using scenario analysis, internal and external data respectively) and then combined as $w_1 F_{SA}(X) + w_2 F_{int}(X) + (1 - w_1 - w_2)F_{ext}(X)$ with some ad-hoc weights $w_1$, $w_2$ to estimate the overall severity distribution. Often frequency distribution is estimated using internal data only, while the severity is modelled by both internal and external data.

In this paper, we suggest a full credibility theory approach (successfully used in the insurance industry and actuarial sciences for many decades) to estimate frequency and severity distributions of the low frequency large losses in each risk cell by taking into account bank internal data, expert opinions and industry data. We present a model, where the Pareto and Poisson distributions are used for modelling severity and frequency respectively. Our approach to estimate the parameter of the Pareto distribution goes back to Rytgaard (1990). Although, the model might be simple it may be very useful at this stage when the data are very limited and it may also have educational impact. Also, it gives a simple example of a consistent credibility approach for estimating operational risk. The external data are incorporated into the model via the hierarchical credibility method described in Bühlmann and Gisler (2005), Chapter 6. The credibility estimators for the severity and frequency distribution parameters, presented in this paper, are based on the use of the Bühlmann-Straub model, developed in 1970 (see Bühlmann and Gisler (2005), Theorem 4.4).

**Bühlmann-Straub model.** Consider a portfolio of $J$ risks modelled by random variables $Y_{j,k} : k = 1,...,K_j$, $j = 1,...,J$ and denote $\mathbf{Y}_j = (Y_{j,1},...,Y_{j,K_j})$. Note that, $K_j$ may vary between the risks. Assume that, for known weights $w_{j,k}$, the $j$-th risk is characterized by an individual risk profile $\theta_j$, which is itself the realization of a random variable $\Theta_j$, and

a) for all $j$, the $Y_{j,k} : k = 1,...,K_j$ are conditionally (given $\Theta_j$) independent with

$$E[Y_{j,k} | \Theta_j] = \mu(\Theta_j), \ \text{Var}[Y_{j,k} | \Theta_j] = \sigma^2(\Theta_j)/w_{j,k} \ ; \tag{2}$$

b) the pairs $(\Theta_1, \mathbf{Y}_1)$, …, $(\Theta_J, \mathbf{Y}_J)$ are independent;

c) $\Theta_1,…,\Theta_J$ are independent and identically distributed.

Define $\mu_0 = E[\mu(\Theta_j)]$, $\sigma^2 = E[\sigma^2(\Theta_j)]$ and $\tau^2 = \text{Var}[\mu(\Theta_j)]$, $j=1,...,J$. Then the homogeneous credibility estimator of $\mu(\Theta_j)$ is given by

$$\hat{\mu}(\Theta_j) = \alpha_j \bar{Y}_j + (1-\alpha_j)\hat{\mu}_0, \tag{3}$$

where

$$\hat{\mu}_0 = \sum_{j=1}^{J} \frac{\alpha_j}{\alpha_0} \bar{Y}_j \ , \ \bar{Y}_j = \sum_{k=1}^{K_j} \frac{w_{j,k}}{\tilde{w}_j} Y_{j,k} \ , \ \alpha_j = \frac{\tilde{w}_j}{\tilde{w}_j + \sigma^2/\tau^2} \ , \ \alpha_0 = \sum_{j=1}^{J} \alpha_j \ , \ \tilde{w}_j = \sum_{k=1}^{K_j} w_{j,k} \ .$$

Hereafter, "hat" is used to denote the estimator of the true value, e.g. $\hat{\mu}_0$ is the estimator of $\mu_0$. The credibility estimators are linear in the observations. They minimize mean square errors of predictions among all linear combinations of the observations. Usually, the credibility estimators are used to estimate expected number of events or expected loss. However, in general, given that the assumptions of the Bühlmann-Straub model are valid, they can be applied to estimate any square integrable valued random variable $Z$ based on some known random vector $\mathbf{Y}$. For example, the elements of $\mathbf{Y}$ can be the maximum likelihood estimators (as in this paper), transformed data, quantiles, etc.

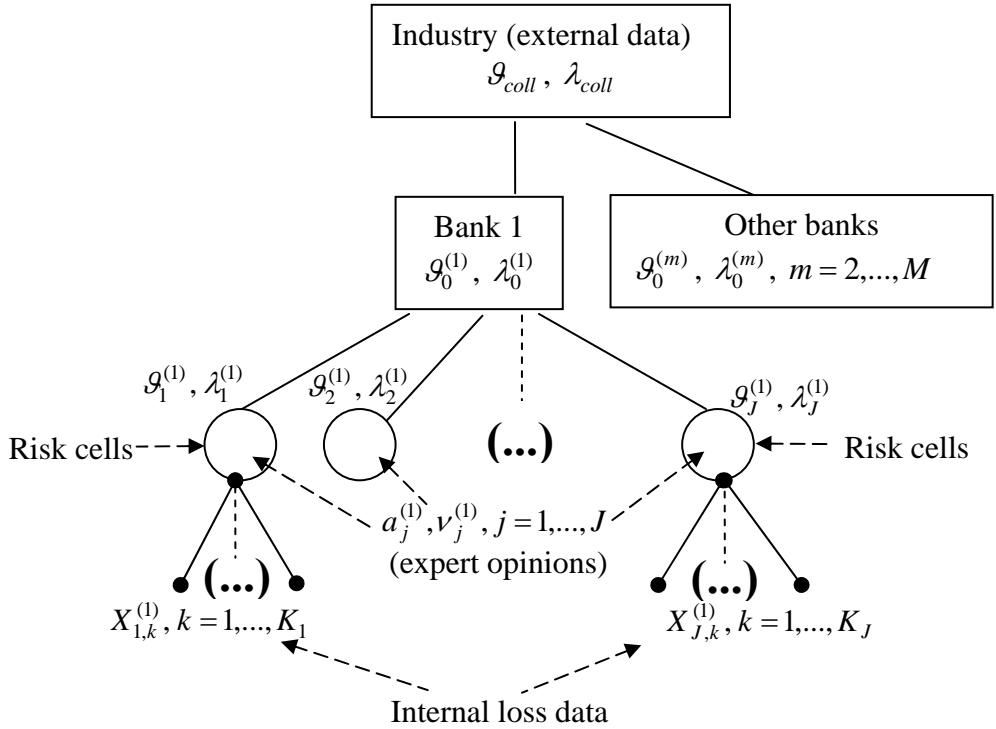

**Figure 1. Main variables and blocks of the "Toy" model for operational risk.** $X_{j,k}^{(1)}$ **are losses (above threshold $L_j^{(1)}$) in risk cells $j = 1,...,J$ of Bank 1. The loss frequency and severity are modelled by the Poisson and Pareto distributions respectively with parameters $\theta_j^{(1)} = v_j^{(1)} \lambda_j^{(1)}$ and $\xi_j^{(1)} = a_j^{(1)} \vartheta_j^{(1)}$ correspondingly. The distribution parameters $\lambda_j^{(1)}$ and $\vartheta_j^{(1)}$ are drawn from common (bank specific) distributions with $E[\lambda_j^{(1)}] = \lambda_0^{(1)}$, $\mathrm{Var}[\lambda_j^{(1)}] = (\omega_0^{(1)})^2$ and $E[\vartheta_j^{(1)}] = \vartheta_0^{(1)}$, $\mathrm{Var}[\vartheta_j^{(1)}] = (\tau_0^{(1)})^2$ respectively. Scaling factors $a_j^{(1)}$ and $v_j^{(1)}$ for the relative differences between the risks can be specified using expert opinions. The structural parameters $\lambda_0^{(m)}$ and $\vartheta_0^{(m)}$ of the banks $m = 1,...,M$ are drawn from common (industry specific) distributions with $E[\lambda_0^{(m)}] = \lambda_{coll}$, $\mathrm{Var}[\lambda_0^{(m)}] = \omega_{coll}^2$ and $E[\vartheta_0^{(m)}] = \vartheta_{coll}$, $\mathrm{Var}[\vartheta_0^{(m)}] = \tau_{coll}^2$ correspondingly.**

## 2 Modelling severity

Consider Bank 1, see Figure 1 (the upper index in all variables is used to refer the bank), and suppose that losses above threshold $L_j^{(1)}$ in the $j$-th risk cell ($j = 1,...,J^{(1)}$) are Pareto distributed with the density and distribution functions given by

$$f(x|\xi_j^{(1)}) = \frac{\xi_j^{(1)}}{L_j^{(1)}}\left(\frac{x}{L_j^{(1)}}\right)^{-\xi_j^{(1)}-1} \text{ and } F(x|\xi_j^{(1)}) = 1 - \left(\frac{x}{L_j^{(1)}}\right)^{-\xi_j^{(1)}} \qquad (4)$$

for $x \geq L_j^{(1)}$, $\xi_j^{(1)} > 0$. It is assumed that $L_j^{(1)}$ are known. This distribution is often used in the insurance industry to model large claims and is a good candidate for modelling large operational risk losses. It is interesting to note that the conditional distribution of the losses exceeding any higher level $\widetilde{L}$ is also a Pareto distribution with parameters $\xi_j^{(1)}$ and $\widetilde{L}$. The results in this section are valid if thresholds are different for different risk cells, although later in the paper, for convenience, we assume the same threshold across risk cells.

Define the Pareto tail parameter as $\xi_j^{(1)} = a_j^{(1)} \vartheta_j^{(1)}$, where $a_j^{(1)}$ are known a priori constants (differences) and $\vartheta_j^{(1)}$ are the risk profiles of the cells in the bank. The constants $a_j^{(1)}$ are scaling factors, reflecting differences in severities across the risks, that can be fixed by experts as discussed below.

**Maximum Likelihood Estimator (MLE) of the tail parameter using data in a risk cell.**
Given that the losses $X_{j,k}^{(1)}$, $k = 1,...,K_j^{(1)}$ in the $j$-th risk cell of Bank 1 are conditionally (given $\vartheta_j^{(1)}$) independent and Pareto distributed, the MLE of $\vartheta_j^{(1)}$ is

$$\hat{\psi}_j^{(1)} = \left[\frac{a_j^{(1)}}{K_j^{(1)}} \sum_{k=1}^{K_j^{(1)}} \ln\left(\frac{X_{j,k}^{(1)}}{L_j^{(1)}}\right)\right]^{-1}. \qquad (5)$$

It is easy to show, see Rytgaard (1990), that an unbiased estimator of $\vartheta_j^{(1)}$ is

$$\hat{\vartheta}_j^{(1)} = \frac{K_j^{(1)} - 1}{K_j^{(1)}} \hat{\psi}_j^{(1)}, \qquad (6)$$

with

$$E[\hat{\vartheta}_j^{(1)} | \vartheta_j^{(1)}] = \vartheta_j^{(1)}, \quad \text{Var}[\hat{\vartheta}_j^{(1)} | \vartheta_j^{(1)}] = (\vartheta_j^{(1)})^2 / (K_j^{(1)} - 2). \qquad (7)$$

A common situation in operational risk is that only a few losses are observed for certain risk cells. Thus, the standard MLE $\hat{\xi}_j^{(1)} = a_j^{(1)} \hat{\vartheta}_j^{(1)}$ (based on the data in the $j$-th risk cell only) for the Pareto parameters $\xi_j^{(1)}$ will not be reliable (this is easy to see from the variance in (7)). The idea is to use the bank's collective losses, industry data and expert opinions to improve the estimates of the Pareto parameters in the risk cells.

**Improved credibility estimator of the tail parameter using all data in the bank.**
The tail parameter estimator $\hat{\xi}_j^{(1)} = a_j^{(1)} \hat{\vartheta}_j^{(1)}$ can be improved using all data in the bank as follows. Assume that, $\vartheta_j^{(1)}$ are independent identically distributed random variables with $E[\vartheta_j^{(1)}] = \vartheta_0^{(1)}$ and $\text{Var}[\vartheta_j^{(1)}] = (\tau_0^{(1)})^2$, where $\vartheta_0^{(1)}$ is a risk profile for the whole bank. Observe that, the unbiased estimators $\hat{\vartheta}_j^{(1)}$, see (7), satisfy the assumptions of the Bühlmann-Straub model (2)-(3) and thus the credibility estimator is given by

$$\hat{\hat{\vartheta}}_j^{(1)} = \alpha_j^{(1)} \hat{\vartheta}_j^{(1)} + (1 - \alpha_j^{(1)}) \vartheta_0^{(1)}, \text{ where } \alpha_j^{(1)} = \frac{K_j^{(1)} - 2}{K_j^{(1)} - 1 + (\vartheta_0^{(1)} / \tau_0^{(1)})^2}. \tag{8}$$

The structural parameters $\vartheta_0^{(1)}$ and $(\tau_0^{(1)})^2$ can be estimated using data across all risk cells in the bank by solving two nonlinear equations (using e.g. iterative procedure, see Bühlmann and Gisler (2005), p.116-117):

$$(\hat{\tau}_0^{(1)})^2 = \frac{1}{J^{(1)} - 1} \sum_{j=1}^{J^{(1)}} \alpha_j^{(1)} (\hat{\vartheta}_j^{(1)} - \hat{\vartheta}_0^{(1)})^2,$$

$$\hat{\vartheta}_0^{(1)} = \frac{1}{W^{(1)}} \sum_{j=1}^{J^{(1)}} \alpha_j^{(1)} \hat{\vartheta}_j^{(1)}, \quad W^{(1)} = \sum_{j=1}^{J^{(1)}} \alpha_j^{(1)}, \tag{9}$$

where coefficients $\alpha_j^{(1)}$ are given in (8), with $\vartheta_0^{(1)}$ and $(\tau_0^{(1)})^2$ replaced by $\hat{\vartheta}_0^{(1)}$ and $(\hat{\tau}_0^{(1)})^2$ respectively. If the solution for $(\hat{\tau}_0^{(1)})^2$ is negative, then we set $\alpha_j^{(1)} = 0$ and $\hat{\vartheta}_0^{(1)} = \frac{1}{W^{(1)}} \sum_{j=1}^{J^{(1)}} w_j^{(1)} \hat{\vartheta}_j^{(1)}$, where $w_j^{(1)} = K_j^{(1)} - 2$ and $W^{(1)} = \sum_{j=1}^{J^{(1)}} w_j^{(1)}$.

The best credibility estimate for the tail parameter in the *j*-th cell (based on the cell data and all data in the bank) is $\hat{\hat{\xi}}_j^{(1)} = a_j^{(1)} \hat{\hat{\vartheta}}_j^{(1)}$. We assumed that constants $a_j^{(1)}$ are known a priori. Note that these constants are defined up to a constant factor, i.e. coefficients $\alpha_j^{(1)}$ (and final estimates of tail parameters) will not change if all $a_j^{(1)}, j = 1,...,J^{(1)}$, are changed/scaled by the same factor. Hence, only relative differences between risks play a role. These constants have the interpretation of a priori differences and can be fixed by expert using opinions on, for example, quantiles of losses exceeding $L_j^{(1)}$. For example, the expert may estimate the probability $q_j$, that the loss in the *j*-th cell will exceed level $T_j$, as $\hat{q}_j$ and use relations $a_j^{(1)} \vartheta_j^{(1)} = -\ln q_j / \ln(T_j / L_j^{(1)})$ and $E[\vartheta_j^{(1)}] = \vartheta_0^{(1)}$ to estimate $a_j^{(1)}$ as $-\ln \hat{q}_j / [\vartheta_0^{(1)} \ln(T_j / L_j^{(1)})]$. Only relative differences play a role so, here (without loss of generality) $\vartheta_0^{(1)}$ can be set equal to 1. Experts may specify several quantiles, then $a_j^{(1)}$ can be estimated using, for example, a least square method. Ideally, the expert specifying constants $a_j^{(1)}$ has a complete overview over all risk cells in the bank, as only relative differences between risks are important. However, in practice, opinions from experts with special knowledge of business specifics within a risk cell are required. Combining opinions from different experts is one of the problems to be resolved by a practitioner. For a more detailed description on using expert opinions for quantification of frequency and severity distributions, see for example Alderweireld, Garcia and Léonard (2006), Shevchenko and Wüthrich (2006).

**Improved credibility estimator of the tail parameter using industry data.**
External data (above threshold $L_j^{(1)}$) can be used to improve the estimate of the tail parameter $\hat{\hat{\xi}}_j^{(1)} = a_j^{(1)} \hat{\hat{\vartheta}}_j^{(1)}$. This can be done using a hierarchical credibility model, Bühlmann and Gisler (2005), Chapter 6. Consider $M$ banks with risk profiles $\vartheta_0^{(m)}$, $m = 1,...,M$. Assume that, $\vartheta_0^{(m)}$ are independent and identically distributed random variables with $E[\vartheta_0^{(m)}] = \vartheta_{coll}$ and $\mathrm{Var}[\vartheta_0^{(m)}] = \tau_{coll}^2$. Then the following statistics and credibility weights are calculated bottom up from the risk cells to industry level.

- Calculate the risk profile estimators $\hat{\vartheta}_j^{(m)}$ and credibility weights $\alpha_j^{(m)}$ for all risk cells $j = 1,...,J^{(m)}$ in the banks $m = 1,...,M$:

$$\hat{\vartheta}_j^{(m)} = \left[ \frac{a_j^{(m)}}{K_j^{(m)} - 1} \sum_{k=1}^{K_j^{(m)}} \ln\left( \frac{X_{j,k}^{(m)}}{L_j^{(m)}} \right) \right]^{-1}, \quad \alpha_j^{(m)} = \frac{K_j^{(m)} - 2}{K_j^{(m)} - 1 + (\vartheta_0^{(m)} / \tau_0^{(m)})^2}. \quad (10)$$

- Calculate the risk profile estimators $\hat{\vartheta}_0^{(m)}$ and credibility weights $\beta^{(m)}$ of the banks $m = 1,...,M$:

$$\hat{\vartheta}_0^{(m)} = \frac{1}{W^{(m)}} \sum_{j=1}^{J^{(m)}} \alpha_j^{(m)} \hat{\vartheta}_j^{(m)}, \quad \beta^{(m)} = \frac{W^{(m)}}{W^{(m)} + (\tau_0^{(m)} / \tau_{coll})^2}, \quad W^{(m)} = \sum_{j=1}^{J^{(m)}} \alpha_j^{(m)}. \quad (11)$$

- Calculate the risk profile estimator $\hat{\vartheta}_{coll}$ for the industry

$$\hat{\vartheta}_{coll} = \frac{1}{A} \sum_{j=1}^{J^{(m)}} \beta^{(m)} \hat{\vartheta}_0^{(m)}, \quad A = \sum_{m=1}^{M} \beta^{(m)}. \quad (12)$$

The final credibility estimators at bank level and in the risk cells are then calculated top-down from the industry level to the individual risk cells:

$$\hat{\hat{\vartheta}}_0^{(1)} = \beta^{(1)}\hat{\vartheta}_0^{(1)} + (1-\beta^{(1)})\hat{\vartheta}_{coll},$$
$$\hat{\hat{\vartheta}}_j^{(1)} = \alpha_j^{(1)}\hat{\vartheta}_j^{(1)} + (1-\alpha_j^{(1)})\hat{\hat{\vartheta}}_0^{(1)}, \; j=1,...,J^{(m)}. \qquad (13)$$

Here, the estimator $\hat{\hat{\vartheta}}_j^{(1)}$ (for the tail parameter in the $j$-th risk cell) is improved, when compared with (8), through improved estimator of $\vartheta_0^{(1)}$. Parameters $\vartheta_0^{(m)}$ and $\tau_0^{(m)}$ can be estimated by solving two equations

$$(\hat{\tau}_0^{(m)})^2 = \frac{1}{J^{(m)}-1}\sum_{j=1}^{J^{(m)}}\alpha_j^{(m)}(\hat{\vartheta}_j^{(m)} - \hat{\vartheta}_0^{(m)})^2,$$
$$\hat{\vartheta}_0^{(m)} = \frac{1}{W^{(m)}}\sum_{j=1}^{J^{(m)}}\alpha_j^{(m)}\hat{\vartheta}_j^{(m)}, \qquad (14)$$

where coefficients $\alpha_j^{(m)}$ and $W^{(m)}$ are given in (10)-(11) with $\vartheta_0^{(m)}$ and $\tau_0^{(m)}$ replaced by their estimators $\hat{\vartheta}_0^{(m)}$ and $\hat{\tau}_0^{(m)}$ respectively. Finally, the parameter $\tau_{coll}$ can be estimated as

$$\hat{\tau}_{coll}^2 = \max\left[c \times \left\{\frac{M}{M-1}\sum_{m=1}^{M}\frac{W^{(m)}}{W_0}(\hat{\vartheta}_0^{(m)} - \overline{\hat{\vartheta}_0^{(m)}})^2 - \frac{M\hat{\tau}^2}{W_0}\right\}, 0\right], \qquad (15)$$

where

$$\hat{\tau}^2 = \frac{1}{M}\sum_{m=1}^{M}(\hat{\tau}_0^{(m)})^2, \; W_0 = \sum_{m=1}^{M}W^{(m)}, \; \overline{\hat{\vartheta}_0^{(m)}} = \frac{1}{M}\sum_{m=1}^{M}\hat{\vartheta}_0^{(m)},$$
$$c = \frac{M-1}{M}\left\{\sum_{m=1}^{M}\frac{W^{(m)}}{W_0}\left(1-\frac{W^{(m)}}{W_0}\right)\right\}^{-1}.$$

**Numerical example**.
To illustrate the above procedures consider an example where losses (exceeding $1million) observed in the Bank 1 with 10 risk cells are given in Table 1 and all risk cells are the same a priori, $a_j^{(1)} = 1, j = 1,...,10$. Using these losses the MLEs for the tail parameters $\hat{\vartheta}_j^{(1)}$, presented in Table 1, are calculated by (6). Then, using (8) and (9), we estimate the bank structural parameters $(\hat{\tau}_0^{(1)})^2 \approx 1.116$ and $\hat{\vartheta}_0^{(1)} \approx 3.157$, and credibility coefficients $\alpha_j^{(1)} \approx 0.446$ (the coefficients are the same because equal number of losses is observed in the cells). The credibility estimators $\hat{\vartheta}_j^{(1)}$, shown in Table 1, are calculated using (8) disregarding industry data. Assume that, given losses across all banks, the industry parameters are estimated as $\hat{\vartheta}_{coll} = 5.0$ and $\hat{\tau}_{coll}^2 = 0.9$. Using (11), we calculate the estimator for the bank credibility coefficient $\hat{\beta}^{(1)} \approx 0.782$. Then formula (13) gives the

estimator for the bank structural parameter $\hat{\hat{\vartheta}}_0^{(1)} \approx 3.558$ improved by industry data. Finally, the credibility estimators $\hat{\hat{\vartheta}}_j^{(1)}$ utilizing industry data are calculated by (13) and presented in Table 1. In this example, the MLEs are quite volatile as the number of observations is small. For example: cell 7 has no large losses and thus its MLE is high; cell 10 has one large loss and thus its MLE is smaller, etc. One could easily calculate cell MLEs vs the number of observations in a cell and observe that MLEs are highly volatile for small number of observations. One large observation may lead to a substantial change in MLE. The credibility estimators (based on data in the bank) are smoother in compare to MLEs. This is because a credibility estimator is a weighted average, according to credibility theory, between a risk cell MLE and the estimator of the bank structural parameter $\hat{\vartheta}_0^{(1)}$ based on all data in the bank. The credibility weights $\alpha_j^{(1)}$ are approximately 0.45 which means that a risk cell MLE (based on observations in a cell) $\hat{\vartheta}_j^{(1)}$ and the a priori estimate $\hat{\vartheta}_0^{(1)} \approx 3.157$ are weighted with 0.45 and 0.55 respectively. Taking into account industry data increases the estimator of the bank structural parameter from $\hat{\vartheta}_0^{(1)} \approx 3.157$ to $\hat{\hat{\vartheta}}_0^{(1)} \approx 3.558$ (because assumed industry collective estimator $\hat{\vartheta}_{coll} = 5.0$ is larger than $\hat{\vartheta}_0^{(1)}$ and the credibility weight $\beta^{(1)}$ for the bank is approximately 0.78). This leads to an increase in the credibility estimators across all cells when industry data are taken into account.

**Table 1: Losses (in millions $) exceeding $1million observed in the Bank 1 and corresponding maximum likelihood and credibility estimators for the Pareto tail parameter in the risk cells.**

| cell 1 | cell 2 | cell 3 | cell 4 | cell 5 | cell 6 | cell 7 | cell 8 | cell 9 | cell 10 |
|---|---|---|---|---|---|---|---|---|---|
| losses (in millions $) exceeding $1million observed in risk cells | | | | | | | | | |
| 1.557 | 9.039 | 1.166 | 1.548 | 1.578 | 1.201 | 1.006 | 1.741 | 1.364 | 1.074 |
| 1.079 | 2.138 | 1.037 | 1.040 | 1.282 | 2.815 | 1.169 | 1.165 | 2.036 | 1.103 |
| 1.047 | 1.008 | 1.136 | 1.045 | 1.092 | 3.037 | 1.215 | 1.010 | 1.014 | 1.664 |
| 1.199 | 1.761 | 2.104 | 1.774 | 1.658 | 1.001 | 1.116 | 1.096 | 1.217 | 1.049 |
| 1.395 | 1.654 | 1.774 | 1.045 | 2.025 | 1.114 | 1.010 | 1.060 | 1.202 | 1.104 |
| 1.060 | 1.073 | 1.161 | 1.856 | 1.129 | 1.422 | 1.560 | 1.352 | 1.095 | 2.924 |
| 3.343 | 2.435 | 1.080 | 1.636 | 1.946 | 2.397 | 1.059 | 1.044 | 1.348 | 1.265 |
| 2.297 | 4.357 | 1.154 | 1.403 | 1.831 | 1.241 | 1.059 | 1.678 | 1.191 | 1.333 |
| 1.297 | 1.576 | 1.257 | 2.522 | 1.478 | 1.522 | 1.050 | 1.882 | 1.161 | 1.424 |
| 1.180 | 1.113 | 1.231 | 1.113 | 1.208 | 1.243 | 1.231 | 1.401 | 1.017 | 1.435 |
| maximum likelihood estimators (MLEs) $\hat{\vartheta}_j^{(1)}$, $j=1,...,10$ | | | | | | | | | |
| 2.499 | 1.280 | 3.688 | 2.487 | 2.264 | 1.992 | 6.963 | 3.335 | 4.194 | 2.870 |
| credibility estimators $\hat{\vartheta}_j^{(1)}$, $j=1,...,10$ disregarding industry data | | | | | | | | | |
| 2.863 | 2.319 | 3.394 | 2.858 | 2.759 | 2.637 | 4.855 | 3.236 | 3.620 | 3.029 |
| credibility estimators $\hat{\hat{\vartheta}}_j^{(1)}$, $j=1,...,10$ utilizing industry data | | | | | | | | | |
| 3.085 | 2.541 | 3.616 | 3.080 | 2.981 | 2.859 | 5.077 | 3.458 | 3.842 | 3.251 |

## 3 Modelling frequency

Again, consider Bank 1, see Figure 1, and the upper index in all variables is used to refer the bank. Let $N_{j,k}^{(1)}$ be the annual number of loss events, exceeding threshold $L_j^{(1)}$, in the $j$-th risk cell ($j = 1,...,J^{(1)}$) in the $k$-th year. Hereafter, for convenience, we assume the same threshold $L$ across all cells and all banks (e.g. one can choose the threshold equal to the threshold in the database of external data). If all data used for estimation are given on the level of risk cells then it is not difficult to modify the model allowing for cell specific thresholds. Assume that $N_{j,k}^{(1)}$ are Poisson distributed with the density function

$$\Pr(N_{j,k}^{(1)} = n | \theta_j^{(1)}) = \frac{(\theta_j^{(1)})^n}{n!} \times \exp(-\theta_j^{(1)}) \tag{16}$$

and moments

$$E[N_{j,k}^{(1)} | \theta_j^{(1)}] = \theta_j^{(1)}, \quad \mathrm{Var}[N_{j,k}^{(1)} | \theta_j^{(1)}] = \theta_j^{(1)}. \tag{17}$$

The arrival rate parameter is defined as $\theta_j^{(1)} = v_j^{(1)} \lambda_j^{(1)}$, where $v_j^{(1)}$ are the known a priori constants and $\lambda_j^{(1)}$ are the risk profiles of the bank cells. The constants $v_j^{(1)}$ are scaling factors, reflecting differences in frequencies across the risks, discussed below.

**Maximum Likelihood Estimator (MLE) of the arrival rate using data in a risk cell.**
Assume that $N_{j,k}^{(1)}$, $k = 1,...,K_j^{(1)}$ in the $j$-th risk cell are conditionally independent (given $\lambda_j^{(1)}$). Then the standard MLE of $\lambda_j^{(1)}$ is

$$\hat{\lambda}_j^{(1)} = \frac{1}{\tilde{v}_j^{(1)}} \sum_{k=1}^{K_j^{(1)}} N_{j,k}^{(1)}, \quad \tilde{v}_j^{(1)} = v_j^{(1)} K_j^{(1)} \tag{18}$$

with

$$E[\hat{\lambda}_j^{(1)} | \lambda_j^{(1)}] = \lambda_j^{(1)}, \quad \mathrm{Var}[\hat{\lambda}_j^{(1)} | \lambda_j^{(1)}] = \lambda_j^{(1)} / \tilde{v}_j^{(1)}. \tag{19}$$

Again, a common situation in operational risk is that only few large losses are observed for some risk cells, so the standard MLEs of parameters $\lambda_j^{(1)}$ will not be reliable. The idea is to use bank collective losses, industry data and expert opinions to improve the estimates of the arrival rate parameters for the risk cells in the bank.

**Credibility estimator of the arrival rate using all data in the bank.**
Assume that $\lambda_j^{(1)}$ are independent identically distributed with $E[\lambda_j^{(1)}] = \lambda_0^{(1)}$ and $\mathrm{Var}[\lambda_j^{(1)}] = (\omega_0^{(1)})^2$. Observe that, the standardized frequencies $F_{j,k}^{(1)} = N_{j,k}^{(1)} / v_j^{(1)}$ satisfy

$$E[F_{j,k}^{(1)} \mid \lambda_j^{(1)}] = \lambda_j^{(1)} \text{ and } \text{Var}[F_{j,k}^{(1)} \mid \lambda_j^{(1)}] = \lambda_j^{(1)} / \nu_j^{(1)}. \tag{20}$$

Thus, $F_{j,k}^{(1)}$ satisfy the Bühlmann-Straub model (2)-(3) and the credibility estimator for $\lambda_j^{(1)}$ is given by

$$\hat{\hat{\lambda}}_j^{(1)} = \gamma_j^{(1)} \hat{\lambda}_j^{(1)} + (1 - \gamma_j^{(1)}) \lambda_0^{(1)}, \text{ where } \gamma_j^{(1)} = \frac{\tilde{\nu}_j^{(1)}}{\tilde{\nu}_j^{(1)} + \lambda_0^{(1)} / (\omega_0^{(1)})^2}. \tag{21}$$

The structural parameters $\lambda_0^{(1)}$ and $\omega_0^{(1)}$ can be estimated using all data in the bank by solving two nonlinear equations (using e.g. iterative procedure, see Bühlmann and Gisler (2005), p.102-103):

$$(\hat{\omega}_0^{(1)})^2 = \max\left[c \times \left\{T - \frac{J^{(1)} \hat{\lambda}_0^{(1)}}{\nu_0^{(1)}}\right\}, 0\right], \quad \hat{\lambda}_0^{(1)} = \frac{1}{\tilde{\gamma}} \sum_j \gamma_j^{(1)} \hat{\lambda}_j^{(1)}, \tag{22}$$

where

$$\nu_0^{(1)} = \sum_{j=1}^{J^{(1)}} \tilde{\nu}_j^{(1)}, \quad T = \frac{J^{(1)}}{J^{(1)} - 1} \sum_{j=1}^{J^{(1)}} \frac{\tilde{\nu}_j^{(1)}}{\nu_0^{(1)}} (\hat{\lambda}_j^{(1)} - \overline{F})^2, \quad \tilde{\gamma} = \sum_j \gamma_j^{(1)},$$

$$\overline{F} = \frac{1}{J^{(1)}} \sum_{j=1}^{J^{(1)}} \hat{\lambda}_j^{(1)}, \quad c = \frac{J^{(1)}}{J^{(1)} - 1} \left\{\sum_{j=1}^{J^{(1)}} \frac{\tilde{\nu}_j^{(1)}}{\nu_0^{(1)}} \left(1 - \frac{\tilde{\nu}_j^{(1)}}{\nu_0^{(1)}}\right)\right\}^{-1}.$$

Here, the coefficients $\gamma_j^{(1)}$ are given in (21) with $\lambda_0^{(1)}$ and $\omega_0^{(1)}$ replaced by $\hat{\lambda}_0^{(1)}$ and $\hat{\omega}_0^{(1)}$ respectively.

The best credibility estimate for the arrival rate parameter in the *j*-th cell (based on the cell data and all data in the bank) is $\hat{\hat{\theta}}_j^{(1)} = \nu_j^{(1)} \hat{\hat{\lambda}}_j^{(1)}$. We assumed that the constants $\nu_j^{(1)}$ are known a priori. Note that these constants are defined up to a constant factor, i.e. the coefficients $\gamma_j^{(1)}$ (and the final estimates of the arrival rate parameters) will not change if all $\nu_j^{(1)}$ are changed by the same factor. Hence, only relative differences between risks play a role. These constants have the interpretation of a priori differences and can be fixed by the expert opinions on expected annual number of losses exceeding threshold *L* for each risk cell. For example, the expert may estimate the expected annual number of events (exceeding threshold *L*) $n_j$ in the *j*-th cell as $\hat{n}_j$ and use relations $\nu_j^{(1)} \lambda_j^{(1)} = n_j$ and $E[\lambda_j^{(1)}] = \lambda_0^{(1)}$ to estimate $\nu_j^{(1)}$ as $\hat{n}_j / \lambda_0^{(1)}$. Only relative differences play a role so, here (without loss of generality) $\lambda_0^{(1)}$ can be set equal to 1. For an example of using expert opinions for quantification of frequency and severity distributions, see Alderweireld, Garcia and Léonard (2006), Shevchenko and Wüthrich (2006).

**Improved credibility estimator of the arrival rate using industry data.**
External data (above threshold $L$) can improve the estimate of arrival rate $\theta_j^{(1)} = v_j^{(1)} \lambda_j^{(1)}$. This can be done using a hierarchical credibility similar to that in the previous section. Below, we present formulas for completeness. Consider $M$ banks with the risk profiles $\lambda_0^{(m)}$, $m=1,...,M$. It is not difficult to introduce a priori differences between the banks but here, for convenience, we assume that, $\lambda_0^{(m)}$ are independent and identically distributed random variables with $E[\lambda_0^{(m)}] = \lambda_{coll}$ and $\mathrm{Var}[\lambda_0^{(m)}] = \omega_{coll}^2$. Then the following statistics and credibility weights are calculated bottom up from the risk cells to the industry level.

- Calculate the risk profile estimators $\hat{\lambda}_j^{(m)}$ and credibility weights $\gamma_j^{(m)}$ for all risk cells $j=1,...,J^{(m)}$ in the banks $m=1,...,M$:

$$\hat{\lambda}_j^{(m)} = \frac{1}{\tilde{v}_j^{(m)}} \sum_{k=1}^{K_j^{(m)}} N_{j,k}^{(m)}, \quad \gamma_j^{(m)} = \frac{\tilde{v}_j^{(m)}}{\tilde{v}_j^{(m)} + \lambda_0^{(m)}/(\omega_0^{(m)})^2}, \quad \tilde{v}_j^{(m)} = v_j^{(m)} K_j^{(m)}. \quad (23)$$

- Calculate the risk profile estimators $\hat{\lambda}_0^{(m)}$ and credibility weights $\rho^{(m)}$ of the banks $m=1,...,M$:

$$\hat{\lambda}_0^{(m)} = \frac{1}{W^{(m)}} \sum_{j=1}^{J^{(m)}} \gamma_j^{(m)} \hat{\lambda}_j^{(m)}, \quad \rho^{(m)} = \frac{W^{(m)}}{W^{(m)} + (\omega_0^{(m)}/\omega_{coll})^2}, \quad W^{(m)} = \sum_{j=1}^{J^{(m)}} \gamma_j^{(m)}. \quad (24)$$

- Calculate the risk profile estimator $\hat{\lambda}_{coll}$ of the industry

$$\hat{\lambda}_{coll} = \frac{1}{A} \sum_{j=1}^{J^{(m)}} \rho^{(m)} \hat{\lambda}_0^{(m)}, \quad A = \sum_{m=1}^{M} \rho^{(m)}. \quad (25)$$

The final credibility estimators at bank level and in the risk cells are calculated top-down as

$$\hat{\hat{\lambda}}_0^{(1)} = \rho^{(1)} \hat{\lambda}_0^{(1)} + (1 - \rho^{(1)}) \hat{\lambda}_{coll},$$
$$\hat{\hat{\lambda}}_j^{(1)} = \gamma_j^{(1)} \hat{\lambda}_j^{(1)} + (1 - \gamma_j^{(1)}) \hat{\hat{\lambda}}_0^{(1)}, \quad j=1,...,J^{(m)}. \quad (26)$$

Here, the estimator $\hat{\hat{\lambda}}_j^{(1)}$ is improved, when compared with (21), through improved estimator of $\lambda_0^{(1)}$. Parameters $\lambda_0^{(m)}$ and $\omega_0^{(m)}$ can be estimated by solving two nonlinear equations (using e.g. iterative procedure) for each bank $m=1,...,M$:

$$(\hat{\omega}_0^{(m)})^2 = \max\left[c \times \left\{T - \frac{J^{(m)} \hat{\lambda}_0^{(m)}}{v_0^{(m)}}\right\}, 0\right], \quad \hat{\lambda}_0^{(m)} = \frac{1}{\tilde{\gamma}} \sum_j \gamma_j^{(m)} \hat{\lambda}_j^{(m)}, \quad (27)$$

where

$$v_0^{(m)} = \sum_{j=1}^{J^{(1)}} \tilde{v}_j^{(m)}, \quad T = \frac{J^{(m)}}{J^{(m)}-1} \sum_{j=1}^{J^{(m)}} \frac{\tilde{v}_j^{(m)}}{v_0^{(m)}} (\hat{\lambda}_j^{(m)} - \overline{F})^2, \quad \tilde{\gamma} = \sum_j \gamma_j^{(m)},$$

$$\overline{F} = \frac{1}{J^{(m)}} \sum_{j=1}^{J^{(m)}} \hat{\lambda}_j^{(m)}, \quad c = \frac{J^{(m)}}{J^{(m)}-1} \left\{ \sum_{j=1}^{J^{(m)}} \frac{\tilde{v}_j^{(m)}}{v_0^{(m)}} \left(1 - \frac{\tilde{v}_j^{(m)}}{v_0^{(m)}}\right) \right\}^{-1}.$$

Here, the coefficients $\gamma_j^{(m)}$ are given in (23) with $\lambda_0^{(m)}$ and $\omega_0^{(m)}$ replaced by $\hat{\lambda}_0^{(m)}$ and $\hat{\omega}_0^{(m)}$ respectively. Finally, the parameter $\omega_{coll}$ can be estimated as

$$\hat{\omega}_{coll}^2 = \max\left[ c \times \left\{ \frac{M}{M-1} \sum_{m=1}^{M} \frac{W^{(m)}}{W_0} (\hat{\lambda}_0^{(m)} - \overline{\hat{\lambda}_0^{(m)}})^2 - \frac{M\hat{\tau}^2}{W_0} \right\}, 0 \right], \quad (28)$$

where

$$\hat{\tau}^2 = \frac{1}{M} \sum_{m=1}^{M} (\hat{\omega}_0^{(m)})^2, \quad W_0 = \sum_{m=1}^{M} W^{(m)}, \quad \overline{\hat{\lambda}_0^{(m)}} = \frac{1}{M} \sum_{m=1}^{M} \hat{\lambda}_0^{(m)},$$

$$c = \frac{M-1}{M} \left\{ \sum_{m=1}^{M} \frac{W^{(m)}}{W_0} \left(1 - \frac{W^{(m)}}{W_0}\right) \right\}^{-1}$$

and coefficients $W^{(m)}$ are given in (24). Note that the above equation is the same as (15) with substitution $\tau_{coll}^2 \to \omega_{coll}^2$, $\vartheta_0^{(m)} \to \lambda_0^{(m)}$, $\tau_0^{(m)} \to \omega_0^{(m)}$, $\beta^{(m)} \to \rho^{(m)}$.

## 4 Remarks and Interpretation

The credibility formulas (8) and (21), for the severity and frequency estimators, based on a cell and bank data, have a simple interpretation. As the number of observations in the $j$-th cell increases, the larger credibility weights $\alpha_j^{(1)}$ and $\gamma_j^{(1)}$ are assigned to the estimators $\hat{\vartheta}_j^{(1)}$ and $\hat{\lambda}_j^{(1)}$ (based on the cell observations) and the lesser weights are assigned to the estimators $\vartheta_0^{(1)}$ and $\lambda_0^{(1)}$ (based on all observations in the bank) respectively. Also, the larger $\tau_0^{(1)}$ and $\omega_0^{(1)}$ (variance across cells in the bank), the larger weights are assigned to $\hat{\vartheta}_j^{(1)}$ and $\hat{\lambda}_j^{(1)}$ correspondingly.

The credibility formulas (10-13) and (23-26), taking into account industry data, have a simple interpretation too. As the number of observations in the bank increases, the larger credibiity weights $\beta^{(1)}$ and $\rho^{(1)}$ are assigned to the estimator $\hat{\vartheta}_0^{(1)}$ and $\hat{\lambda}_0^{(1)}$ (based on the bank observations) and the lesser weight is assigned to the estimators $\hat{\vartheta}_{coll}$ and $\hat{\lambda}_{coll}$ (based on observations across all banks) respectively. Also, the larger $\tau_{coll}$ and $\omega_{coll}$ (variance across the banks), the larger weights are assigned to $\hat{\vartheta}_0^{(1)}$ and $\hat{\lambda}_0^{(1)}$ correspondingly. For a detailed discussion on the credibility parameters, we refer to Bühlmann and Gisler (2005) Section 4.4.

In a pure Bayesian approach, the industry risk profiles $\vartheta_{coll}, \tau_{coll}$ and $\lambda_{coll}, \omega_{coll}$ would be given a priori (often practice in the insurance industry). Then equations (9) and (22) should be solved for a bank only and estimators (for the tail parameter and arrival rate) in the bank risk cells are calculated using (13) and (26). It would be ideal if the industry risk profiles are calculated and provided by the regulators to ensure consistency across the banks. Also, if the banks would disclose their risk profile information then the industry profiles can be estimated using (11, 12) and (24, 25). Unfortunately this may not be realistic at the moment, then the industry risk profiles should be estimated by complete procedures as described in the above sections, using equations (10-12) and (23-25) for the severity and frequency estimators respectively. It will require knowledge of the large losses (above some large threshold) experienced by other banks available at the moment through external databases. For practical purposes, it is convenient to choose the same threshold $L$ across all risk cells and all banks. One can choose the threshold equal to the threshold in the external database (e.g. $L$=$1Million). Then, vendors data (typically recorded above US$1Million) can be used for calibration of the proposed model. If available, consortium-based data can be used for model calibration (it is expected that the threshold $L$ in the proposed model is larger than the consortium-based data collection threshold). The data quality and survival biases in external databases are the issues that should be considered in practice but go beyond the purposes of this paper.

## 5 The capital calculations

For the purposes of the regulatory capital calculations of operational risk, the annual loss distribution, in particular its 0.999 quantile (VaR) as a risk measure, should be quantified for each Basel II risk cell in the matrix of eight business lines times seven risk types and for the whole bank. The credibility model presented in this paper is proposed for modeling low frequency high impact losses exceeding some large threshold $L$. Of course, modelling of the high frequency low impact losses (below threshold $L$) should be added to the model before the final operational risk capital charge is estimated (for a related actuarial literature on this topic see Sandström (2006) and Wüthrich (2006)). That is, we suggest that the losses above threshold $L$ are modelled using credibility theory as described in this paper, while the losses below the threshold are modelled separately. Note that typically, the low frequency high impact losses give largest contribution to the final capital charge. The number of high frequency low impact losses recorded in the bank internally is usually large enough to obtain reliable estimates by a standard fitting of the frequency and severity distributions without the use of the external data. Modelling these losses is important but goes beyond the purposes of this paper.

The total LDA model for an annual loss $Z_j$ in risk cell $j$ (combining losses below and above $L$) is

$$Z_j = \sum_{n=1}^{N_j^{(lf)}} X_{j,n}^{(lf)} + \sum_{k=1}^{N_j^{(hf)}} X_{j,k}^{(hf)}. \tag{29}$$

Here, $N_j^{(lf)}$ and $X_{j,n}^{(lf)}$, $n=1,...,N_j^{(lf)}$ are the annual number of events and iid severities of the low frequency high impact losses (exceeding $L$) modelled by distributions $P_j^{(lf)}(.)$ and $F_j^{(lf)}(.)$ respectively. $N_j^{(hf)}$ and $X_{j,k}^{(hf)}$, $k=1,...,N_j^{(hf)}$ are the annual number events and iid severities of the high frequency low impact losses (below $L$) modelled by distributions $P_j^{(hf)}(.)$ and $F_j^{(hf)}(.)$ respectively. It is reasonable to assume that the low frequency high impact losses and high frequency low impact losses are independent, i.e: $X_{j,n}$ and $Y_{j,k}$ are independent, $N_j^{(lf)}$ and $N_j^{(hf)}$ are independent. As usual, independence between severities and frequencies is assumed. In the credibility model described in this paper, $P_j^{(lf)}(.)$ and $F_j^{(lf)}(.)$ are the Poisson and Pareto distributions with parameters $\theta_j$ and $\xi_j$ estimated by $v_j^{(1)}\hat{\hat{\lambda}}_j^{(1)}$ and $a_j^{(1)}\hat{\hat{\vartheta}}_j^{(1)}$ respectively as described in the above sections. Then, the annual loss distribution for each risk cell and the whole bank can be calculated using, for example, the Monte Carlo procedure with the following logical steps (where all random samples are independent):

**Step1**. For each risk cell $j=1,...,J$ simulate the annual number of events $N_j^{(lf)}$ and $N_j^{(hf)}$ from the distributions $P_j^{(lf)}(.)$ and $P_j^{(hf)}(.)$ respectively.

**Step2**. Given $N_j^{(lf)}$ and $N_j^{(hf)}$ from the Step 1, simulate severities $X_{j,n}^{(lf)}$, $n=1,...,N_j^{(lf)}$ and $X_{j,k}^{(hf)}$, $k=1,...,N_j^{(hf)}$ from distributions $F_j^{(lf)}(.)$ and $F_j^{(hf)}(.)$ respectively.

**Step3**. Find the annual loss $Z_j$ for each risk cell $j=1,...,J$ using formula (29) and the annual loss in the bank as $Z = \sum_{k=1}^{J} Z_j$.

**Step4**. Repeat Steps 1-3 $K$ times to build samples of the annual losses $Z(k)$ and $Z_j(k)$, $k=1,...,K$. Then, the 0.999 quantile (and other distribution characteristics if required) is estimated using simulated sample in a usual way.

In the above procedure we assumed that risk cells are independent. While it is important (and quite realistic) assumption of the proposed model that the low frequency high impact losses from different risk cells are independent, dependence can be considered between the high frequency low impact losses from different risk cells. The dependence can be introduced using for example copula methods. For further information on the application of the copula method in finance, we refer to McNeil, Frey and Embrechts (2005). Also, see Frachot, Roncalli and Salomon (2004) for a discussion of dependence between operational risks. The modelling of common events (shocks) that affect many risk cells simultaneously is another important part of operational risk modelling. These common shocks can be modelled as separate compound processes and mapped to the risk cells, effectively introducing dependence between risk cells, see Lindskog and McNeil (2003). Accurate quantification of the dependencies between the risks is a difficult task, which is an open field for future research.

The qualitative impact of considering industry data on the bank VaR figures is quite intuitive for the proposed credibility model. If the industry risk profile $\lambda_{coll}$ is higher than the bank profile $\lambda_0^{(1)}$ based on the internal data, then the credibility estimators of the bank profile and arrival rate in risk cells will increase, as follows from (26), and vice versa. Increase in the arrival rate estimators in all risk cells of the bank will increase VaR figures for the bank and vice versa. Also, if the industry risk profile $\vartheta_{coll}$ is larger than the bank profile $\vartheta_0^{(1)}$ based on the internal data, then the credibility estimators for the bank profile and Pareto tail parameter in risk cells will increase, as follows from (13), and vice versa (this behaviour of the credibility estimators is demonstrated in the numerical example, Table 1). Increase of the Pareto tail parameter across all risk cells will decrease VaR figures for the bank and vice versa. The magnitudes of these impacts will depend on credibility weights of the bank $\beta^{(1)}$ and $\rho^{(1)}$ discussed in the previous section.

## 6 Conclusions

In this paper, we proposed the use of a full credibility theory approach to estimate parameters of the frequency (Poisson) and severity (Pareto) distributions for the low frequency high impact operational risk losses exceeding some threshold for each risk cell. The initial estimators are based on the use of risk cell data. Then the estimators are improved by the use of all data in the bank (across all risk cells) and by experts who may specify the relative difference between risks. Finally, the estimators are corrected by the industry data. A numerical example of the procedure for the loss severities is given in Table 1 (the procedure for the loss frequencies is very similar). The main reason for the proposed approach is that, typically, the bank's internal data of the large losses in risk cells are so limited that the standard maximum likelihood estimates are not reliable. The proposed credibility theory approach allows to use the bank's collective losses, external data and expert opinions to improve the estimates.

    The described model is not too complicated but it gives an example of a full credibility approach that we believe is well suited for operational risk quantification. It has a simple structure which is beneficial for practical use and can engage the bank risk managers, statisticians and regulators in productive model development and risk assessment. The model provides a framework that can be developed further by considering other distribution types and dependencies between risks. As it stands, the model does not have a time component and incorporation of evolutionary models, where the risk profiles are evolving in time, is important for further development. Justification of the model assumptions (such as conditional independence between the losses or common distribution for the risk profiles across the risks) can be based on the analysis of the unconditional properties (e.g. unconditional means, covariances) of the losses and should be addressed during model implementation. Adding extra levels to the considered hierarchical structure may be required to model the actual risk cell structure in a bank.

    Of course, a full Bayesian approach would allow for quantification not only better point estimates for the distribution parameters but their whole posterior distributions, see e.g. Shevchenko and Wüthrich (2006). However, it requires specification of the prior distributions that can often be difficult in the case of very limited data. We hope that

practitioners in operational risk will find the presented model useful both from a practical and an educational point of view.

## Acknowledgements
We are very grateful to Paul Embrechts for his support and encouragement.## References
Alderweireld, T., J. Garcia and L. Léonard (2006). A practical operational risk scenario analysis quantification. *Risk Magazine*, February 2006, 93–95.

BIS (2005). *Basel II: International Convergence of Capital Measurement and Capital Standards: a revised framework*. Bank for International Settlements (BIS), www.bis.org

Bühlmann, H. and A. Gisler (2005). *A Course in Credibility Theory and its Applications*. Springer-Verlag, Berlin.

Chavez-Demoulin, V., P. Embrechts and J. Nešlehová (2006). Quantitative Models for Operational Risk: Extremes, Dependence and Aggregation. *Journal of Banking and Finance* 30(10), 2635-2658.

Cruz, M., editor (2004). *Operational Risk Modelling and Analysis: Theory and Practice*. Risk Books, London.

Frachot, A., T. Roncalli and E. Salomon (2004). The Correlation Problem in Operational Risk. Groupe de Recherche Opérationnelle, France. Working paper.

Lindskog, F. and A. McNeil (2003). Common Poisson shock models: Application to insurance and credit risk modelling. *ASTIN Bulletin* 33, 209-238.

McNeil, A. J., R. Frey and P. Embrechts (2005). *Quantitative Risk Management: Concepts, Techniques and Tools*. Princeton University Press.

Rytgaard, M. (1990). Estimation in Pareto distribution. *ASTIN Bulletin* 20, 201-216.

Sandström, A. (2006). *Solvency: Models, Assessment and Regulation*. Chapman & Hall/CRC, Boca Raton.

Shevchenko, P.V. and M.V. Wüthrich (2006). The Structural Modelling of Operational Risk via Bayesian Inference: Combining Loss Data with Expert Opinions. *The Journal of Operational Risk* **1**(3), 3-26.

Wüthrich, M. V. (2006). Premium Liability Risks: Modelling Small Claims. *Bulletin of the Swiss Association of Actuaries*, 1, 27-38.18